\pdfoutput=1

\documentclass[twocolumn,11pt]{article}
\usepackage[authoryear,colon]{natbib}
\usepackage{amsmath}            
\usepackage{amssymb}            
\usepackage{wrapfig}            
\usepackage[ansinew]{inputenc}   
\usepackage{fancyhdr}           
\usepackage{lscape}             
\usepackage{hyperref}           
\usepackage{capt-of}            
\usepackage{makeidx}            
\usepackage{manyfoot,footmisc}
\DefineFNsymbols{stars}[text]{*{**}{***}{****}}
\setfnsymbol{stars} \DeclareNewFootnote{T}[fnsymbol]

\usepackage{geometry}
\usepackage{ifpdf}
\makeatletter

\renewcommand\section{\@startsection {section}{1}{\z@}%
                                   {-3.5ex \@plus -1ex \@minus -.2ex}%
                                   {2.3ex \@plus.2ex}%
                                   {\centering\Large\scshape}}
\makeatletter
\renewcommand\subsection{\@startsection {subsection}{1}{\z@}%
                                   {-3.5ex \@plus -1ex \@minus -.2ex}%
                                   {2.3ex \@plus.2ex}%
                                   {\large}}
\renewcommand\subsubsection{\@startsection {subsubsection}{1}{\z@}%
                                   {-3.5ex \@plus -1ex \@minus -.2ex}%
                                   {2.3ex \@plus.2ex}%
                                   {\hspace*{1em}\large}}

\newcommand{\leiden}{Leiden University}
\newcommand{\strath}{University of Strathclyde}
\newcommand{\tubs}{Technical University at Braunschweig}
\newcommand{\rug}{University of Groningen}

\ifpdf
\usepackage[pdftex,final]{graphicx}
\usepackage{epstopdf}
\else
\usepackage{graphicx}
\fi
\input amssym.def
\input amssym.tex
\hypersetup{pdftitle = {Experimental Studies on the Aggregation
Properties of Ice and Dust in Planet-Forming Regions}, pdfsubject
= {Contribution to the 58th International Astronautical Congress
2007}, pdfauthor = {Daniel Hei�elmann}, pdfstartview=FitH,
bookmarksopen=true, bookmarksnumbered=true}

\begin{document}

\twocolumn[{\csname @twocolumnfalse\endcsname
\begin{center}
\Large
\textsc{Experimental Studies on the Aggregation Properties of Ice
and Dust in Planet-Forming Regions}

\vspace{1cm}\normalsize
Daniel Hei\ss elmann\\
\emph{Institut f\"ur Geophysik und extraterrestrische Physik,
Technische Universit\"at zu Braunschweig,\\
Mendelssohnstra\ss e 3, D-38106 Braunschweig, Germany}\\
\href{mailto:d.heisselmann@tu-bs.de}{\texttt{d.heisselmann@tu-bs.de}},\\
\vspace{2.5mm}
Helen J. Fraser\\
\emph{Department of Physics, University of Strathclyde, John
Anderson Building, 107 Rottenrow, Glasgow G4 0NG, United Kingdom}\\
\href{mailto:h.fraser@phys.strath.ac.uk}{\texttt{h.fraser@phys.strath.ac.uk}},\\
\vspace{2.5mm} and\\ \vspace{2.5mm}
J\"urgen Blum\\
\emph{Institut f\"ur Geophysik und extraterrestrische Physik,
Technische Universit\"at zu Braunschweig,\\
Mendelssohnstra\ss e 3, D-38106 Braunschweig, Germany}\\
\href{mailto:j.blum@tu-bs.de}{ \texttt{j.blum@tu-bs.de}}
\\
\end{center}

\begin{abstract}
\noindent
\rule[4pt]{\linewidth}{1.2pt}\\
To reveal the formation of planetesimals it is of great importance
to understand the collision behavior of the dusty and icy
aggregates they have formed from. We present an experimental setup
to investigate the aggregation properties in low-velocity
collisions of dust aggregates, solid ices and icy aggregates under
microgravity conditions. Results from ESA's $\mathrm{45^{th}}$
Parabolic Flight Campaign show that most collisions in the
velocity range $0.1\,\mathrm{m\,s^{-^1}} \lesssim v_c \lesssim
0.5\,\mathrm{m\,s^{-^1}}$ are dominated by a rebound behavior of
the projectile dust aggregates and only $\sim 5\%$ of the
translational kinetic energy is conserved after the
encounters.\\
\rule[4pt]{\linewidth}{1.2pt}

\end{abstract}

}]
\section{\underline{Introduction}}
Current theories say that stars form from gravitational collapse
of dense, turbulent, and rotating molecular clouds. This
contraction occurs when the gas pressure can no longer support the
cloud core against gravity. After the formation of a central
protostar the gas environment cools down and allows for the
condensation of (sub-)\linebreak micrometer-sized mineral grains.
Due to the conservation of angular momentum dust is accreted along
the rotational axis of the system by the new-born star whereas
perpendicular to this axis a flattened, differentially rotating
accretion disk develops. In the protoplanetary disk, which
typically has a dust-to-gas ratio of $\sim0.01$, matter can only
be transported towards the center of gravity as a result of
friction forces, which are thought to be caused by gas turbulence
or magneto-rotational turbulence. According to the model of
\citet{wei_cuzz1993}, inelastic collisions of dust grains combined
with adhesive surface forces (van-der-Waals forces or hydrogen
bonding) between colliding particles cause them to coagulate by a
hit-and-stick method. The relative velocities of the colliding
grains or agglomerates are the result of Brownian motion
\citep{brown1828PhilJ,einstein1905Brown}, relative drift motion of
two particles or gas turbulence \citep{wei_cuzz1993}. In this
scenario Brownian motion is most important for small dust
aggregates of sizes between $\sim1\,\mathrm{\mu m}$ and
$\sim100\,\mathrm{\mu m}$ while collisions of larger aggregates
are dominated by gas turbulence and drift towards the center of
gravity \citep{wei_cuzz1993}.\par

The so-formed agglomerates become more compact due to
restructuring in more energetic collisions which occur due to the
increasing relative velocities with increasing agglomerates
masses. When the particles have become more compact they decouple
from the protoplanetary disk's gas and sedimentation to the
mid-plane causes the onset of the runaway-growth phase
\citep{wei_cuzz1993,wei1997} in which larger aggregates increase
their size dramatically by sweeping-up smaller agglomerates while
sedimentating.\par

As soon as planetesimals of  $s\gtrsim1\,\mathrm{km}$ have formed
the collisions between those bodies are dominated by gravity.
Gravitational perturbations accelerate the smaller planetesimals
leading to high-velocity collisions between kilometer-sized
bodies. These encounters result in collisional fragmentation from
which the environment of the larger planetesimals is fed with
smaller fragments. Capturing of fragments by larger runaway bodies
then allows the rapid growth to Mercury size on a $\sim10^5$-years
time scale
\citep{wetherill_stewart1989Icarus,wetherill_stewart1993Icarus}.\par
However, the growth of kilometer-sized planetesimals from
millimeter to centimeter-sized aggregates is still an unanswered
question. Sect. \ref{s_prev_work} of this manuscript gives an
overview of the numerous simulations and experimental studies on
the formation of planetesimals which have been carried out within
the past decades. In Sect. \ref{s_objectives} the scientific
objectives of the experiments performed here are described and
Sect. \ref{s_experiment} gives a detailed description of the setup
used for studying low-velocity collisions of dust aggregates. The
results obtained during ESA's $\mathrm{45^{th}}$ Parabolic Flight
Campaign and their interpretation can be found in Sect.
\ref{s_micro_g_exp}. In Sect. \ref{s_improvements} several
intended improvements to the experiment setup and procedures are
explained.\par

\section{\underline{Previous Work}}\label{s_prev_work}
In the past decades numerous laboratory experiments and numerical
simulations were performed to reveal the processes involved in the
formation of the terrestrial planets, the asteroids and the cores of
the giant planets.\par

\citet{wei_cuzz1993} present a model which describes the formation
of planetesimals from fractal dust aggregates which have formed in
the collisions of single, (sub-)\linebreak micrometer-sized dust
grains \citep{meakin_donn1988ApJ,blum2004ASPC,krause_blum2004}.

In numerical simulations \citet{dominik_tielens1997ApJ} studied
the interaction forces of two colliding aggregates of fractal
structure, which consist of equal-sized, spherical monomers. They
considered the adhesion forces, the rolling and the sliding
between all pairs of grains which are in contact inside the two
colliding aggregates. The outcome of their calculations is that
the key parameters in collisions of fractal aggregates are the
number of monomer-monomer contacts within both aggregates, the
energy required to break-up a grain-grain contact and the
rolling-friction energy which is necessary to roll two neighboring
grains by one quarter of their circumference. With increasing
collision velocity and thus with increasing impact energy,
\citeauthor{dominik_tielens1997ApJ} distinguish between five
regimes: (1) a hit-and-stick behavior in which sticking occurs
without restructuring of the aggregates, (2) the beginning of
restructuring resulting in compaction of the aggregates until (3)
a state of maximum compaction is achieved. At even higher impact
velocities (4) the break-up of grain-grain bonds occurs and
finally results in (5) the catastrophic disruption of the
aggregates. \citet{blum_wurm2000Icarus} experimentally showed that
this model is correct and measured at which collision velocities
the various stages occur.\par

\citet{wurm_blum1998} present results from laboratory experiments
probing the collisions of fluffy, fractal aggregates
($D_f\approx1.91$, where $m \propto r^{D_f}$ and $m$, $r$, and
$D_f$ are the mass, the radius of gyration and the fractal
dimension, respectively) consisting of monodisperse spherical
$\mathrm{SiO_2}$ grains ($s_0=1.9\,\mathrm{\mu m}$). They found
that in the collision velocity range of
$0.001\,\mathrm{m\,s^{-1}}\lesssim \mathnormal{v_c} \lesssim
0.01\,\mathrm{m\,s^{-1}}$ the sticking probability is unity and no
restructuring or compaction of the constituent aggregates
occurs.\par

\citet{blum_schraepler2004} and \citet{blum_et_al2006} built
large, compact ($D_f\approx 3$) dust agglomerates from
$1.5\,\mathrm{\mu m}$-sized $\mathrm{SiO_2}$ grains, which can be
considered as planetesimal analogue bodies. These agglomerates --
so-called `dust cakes' -- are produced by the method of Random
Ballistic Deposition (RBD) in which first dust powder is
de-agglomerated to its micrometer-sized constituent grains,
secondly the grains are coupled to a gas flow and finally they are
unidirectionally deposited on a filter. By using this technology
planetesimal-analog aggregates of $2.5\,\mathrm{cm}$ diameter and
up to $2\,\mathrm{cm}$ in height can be built.\par

\citet{blum_schraepler2004} and \citet{blum_et_al2006} showed that
these aggregates have a highly porous structure. In unidirectional
compression experiments the volume filling factor $\phi =
\frac{\varrho}{\varrho _0}$ (where $\varrho$ is the aggregate's
density and $\varrho _0$ is the bulk density of the constituent
monomer grains) as a function of the applied pressure increased
asymptotically from $\phi = 0.07...0.15$ to $\phi\lesssim
0.20...0.33$.\par

\citet{langkowski_et_al2007} recently carried out microgravity
experiments to investigate the collision behavior of $s \sim 0.1 -
1\,\mathrm{mm}$-sized high-porosity dust-aggregates (projectile)
and $2.5\,\mathrm{cm}$-sized target of the same material in the
velocity regime of $0.5 - 3\,\mathrm{m\,s^{-1}}$. They found that
near-normal collisions were dominated by sticking while
near-tangential impacts showed rebounding behavior of the
projectiles. For collisions which did not result in sticking of
the projectile \citeauthor{langkowski_et_al2007} observed a mass
transfer from the target to the projectile which approximately
doubled the projectile's mass.\par

Laboratory collision experiments with pairs of
$1\,\mathrm{mm}$-sized aggregates at velocities of $0.15$ to
$3.9\,\mathrm{m\,s^{-1}}$ were carried out by
\citet{blum_muench1993Icar}. They used aggregates consisting of
$s_0=0.2...1\,\mathrm{\mu m}$-sized $\mathrm{ZrSiO_4}$ monomers
which have a volume filling factor of $\phi=0.26$ and collided
them at arbitrary impact angles. It was found that low-velocity
encounters resulted in rebounding of the projectile aggregates,
whereas at $v_c\gtrsim 1\,\mathrm{m\,s^{-1}}$ a transition to
fragmentation was observed. For the bouncing aggregates the
coefficient of restitution $\epsilon$ -- the ratio of relative
velocities after and before the collision -- and the normalized
translational energy $\epsilon ^2$ were determined. The results
showed that for central collisions the conserved translational
energy was at $\sim 10\%$ of the value before the collision. For
perfectly grazing encounters a theoretical value for the conserved
kinetic energy of $\epsilon ^2_t=0.51$ was calculated.
Furthermore, \citeauthor{blum_muench1993Icar} observed that
collisions at velocities exceeding $\sim 3\,\mathrm{m\,s^{-1}}$
were followed by complete disruption of the initial impactors with
the number of fragments following a power-law mass
distribution.\par

In addition to silicates, water ices (and other condensates of
volatiles, like e.g. $\mathrm{CO,\; CO_2,\; NH_3,\;CH_3OH}$ and
$\mathrm{CH_4}$), which are ubiquitous in the outer reaches of the
solar nebula, are of great importance for the formation of Pluto,
the comets, the Kuiper belt objects (KBOs) and the icy moons of
the giant planets. However, only very limited knowledge exists on
the aggregation properties of ices in the solar nebula and
detailed studies of the collision behavior and fragmentation
thresholds of icy agglomerates
\citep{ehrenfreund2003,fraser_et_al2005ESA}.\par

To probe the collision properties of icy bodies as they are found
in Saturn's rings, \citet{bridges_et_al1984Nature} and
\citet{hatzes_et_al1988} performed quasi-2D collision experiments
between solid ice spheres of several centimeters in diameter and
at very low relative velocities of $1.5\cdot 10^{-4} - 2\cdot
10^{-2}\,\mathrm{m\,s^{-1}}$. They found that the coefficient of
restitution $\epsilon$ follows an exponential law of the type
$\epsilon \left(v_c\right)=C\cdot \exp \left(-\gamma v_c\right)$.
It was also discovered that the values of $\epsilon$ strongly
depend on the properties of the contact surfaces and can be
lowered by up to 30\% as a result of  roughened or frosted
surfaces. \citeauthor{hatzes_et_al1988} conclude that, except for
particles with a very smooth surface, a size dependence exists in
the coefficient of restitution, $\epsilon$.

\section{\underline{Scientific Objectives}}\label{s_objectives}
The goal of our experiments is to analyze if and how
millimeter-sized dust aggregates, solid ices and icy aggregates
stick together (i.e. form aggregates) in a microgravity
environment that best simulates the protoplanetary disks where
these particles combine to form planetesimals and cometary nuclei.
Therefore, we carry out collisions of pairs of
$2-5\,\mathrm{mm}$-sized projectiles probing mutual collisions of
similar sized aggregates in protoplanetary disks. To simulate
collisions of different sized bodies the impacts of aggregates on
a larger solid target can also be observed. In both types of
experiments the collisions occur at low velocities of
$0.1\,\mathrm{m\,s^{-^1}} \lesssim v_c \lesssim
0.5\,\mathrm{m\,s^{-^1}}$ which are the typical relative
velocities of millimeter to centimeter-sized aggregates in the
solar nebula \citep{wei_cuzz1993,wei1997}. The results will
provide insight into current aggregation theories, as well as
increase our knowledge on the initial conditions and clumping
properties leading to the creation of planetary systems.\par

On ESA's 45$\mathrm{^{th}}$ Parabolic Flight Campaign we
investigated the collision behavior of RBD dust aggregates
\citep{blum_schraepler2004,blum_et_al2006} at room temperature
corresponding to a distance of $1\,\mathrm{AU}$ (the distance of
Earth's orbit to the sun) from the protostar \citep{wood2000SSRv}.
In two more Parabolic Flight Campaigns, scheduled for November
2007, we will continue our experiments to probe collisions and
impacts of similar dust aggregates at low temperatures of
$140-250\,\mathrm{K}$ ($2-5\,\mathrm{AU}$) and we will perform
collisions using icy samples at cryogenic temperatures ($\leq
140\,\mathrm{K}$), similar to those in the outer protoplanetary
nebula ($5-30\,\mathrm{AU}$).\par

\section{\underline{Experiment Setup}}\label{s_experiment}

\begin{figure}[t!b]
        \begin{center}
            \includegraphics[width=1.0\columnwidth]{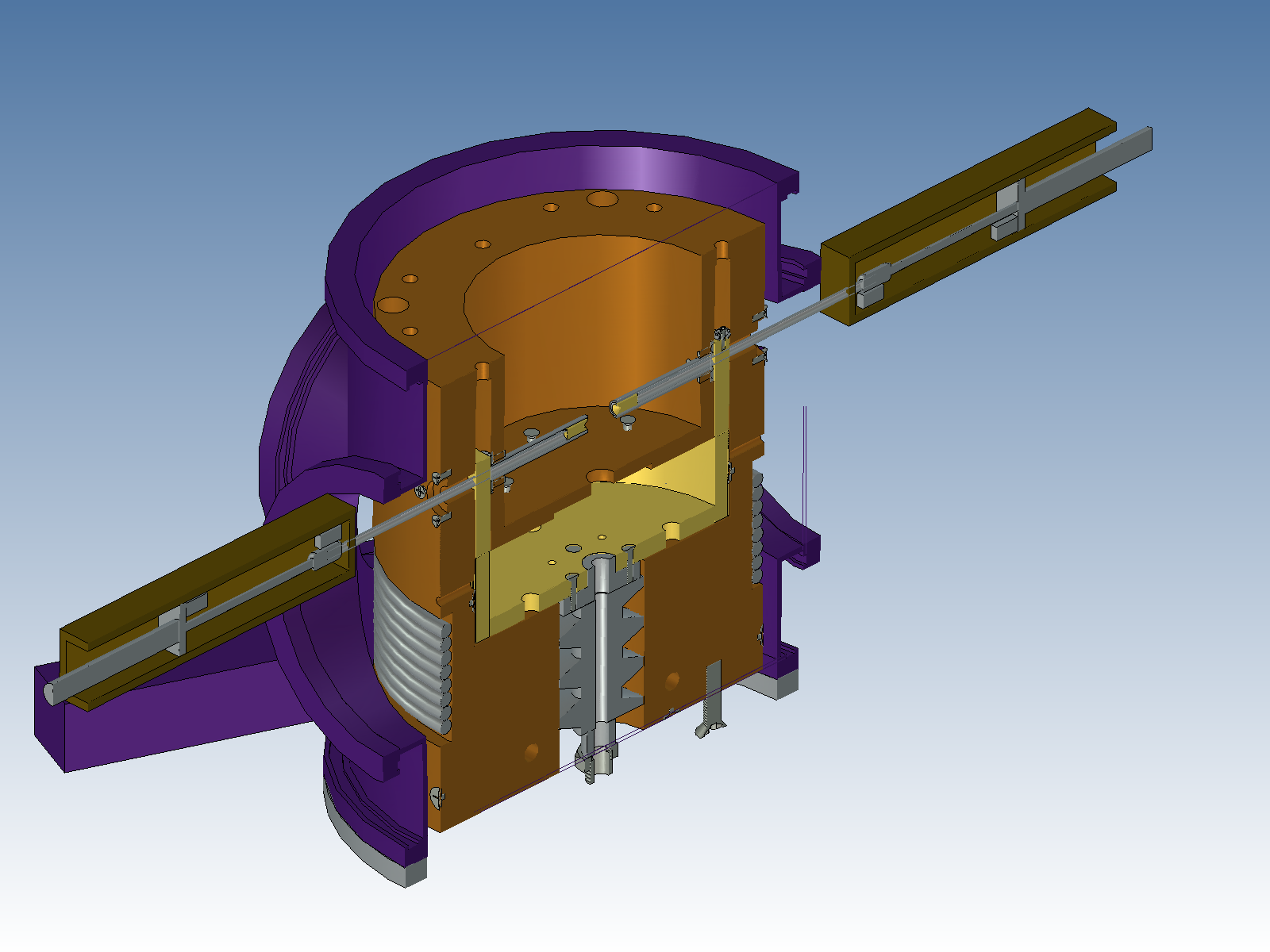}
            \caption{Cross-section of the experiment chamber. At left and right, two hydraulic actuators are attached. In the center,
            the shielded sample repository is fit onto the thermal reservoir.}
            \label{f_CAD}
        \end{center}
\end{figure}

To investigate the collisions of millimeter-sized dust aggregates
and icy bodies under conditions comparable to the early solar
nebula an experimental setup was designed to allow a large number
of collisions at temperatures of $80-300\,\mathrm{K}$. Therefore,
a sample storage for 180 projectiles was fit onto a thermal
reservoir (Sect. \ref{ss_storage}) which keeps the desired
temperatures for the duration of the experiment (see the CAD
drawing in Fig. \ref{f_CAD}). To prevent the fragile dust
aggregates from being damaged by acceleration of more than
$10\,g_0$ (where $g_0=9.81\,\mathrm{m\,s^{-2}}$, Earth's
gravitational acceleration) a smooth particle acceleration (Sect.
\ref{ss_pistons}) mechanism was developed. To probe collisions of
different sized bodies, a target frame -- simulating a larger
aggregate -- is installed in the center of the collision volume
(Sect. \ref{ss_target}). The collision events are recorded by a
high-speed, high-resolution digital recording system (Sect.
\ref{ss_camera}) attached to the vacuum chamber's top flange. The
vacuum chamber (housing the collision volume and the sample
repository) and all diagnostic hardware are installed into two
custom-made aluminum racks to comply with the safety requirements
for parabolic flights.

\subsection{\underline{Thermal Reservoir and Particle}\\\underline{Storage}}\label{ss_storage}

To keep the experiment at the desired low and cryogenic
temperatures for the duration of one parabolic flight a thermal
reservoir is mandatory, since no liquid Nitrogen ($\mathrm{N_2}$)
is allowed aboard the Zero-G aircraft. Therefore, the experiment
is equipped with a massive, $50\,\mathrm{kg}$ copper block which
is fit into a cylindrical vacuum chamber of
$290\,\mathrm{mm}\times 250\,\mathrm{mm}$. The thermal reservoir
is cooled by liquid $\mathrm{N_2}$ spiraling through a copper tube
wound around the copper block and attached to a liquid
$\mathrm{N_2}$ feedthrough at the bottom flange of the vacuum
chamber. Cooled down to cryogenic temperatures the warm-up rate of
the system is $\lesssim5\,\mathrm{K}$ per hour. To avoid heat
transfer by thermal conductivity of the residual air inside the
vacuum chamber a dry, oil-free membrane pump and a turbo molecular
pump (TMP) are operated in series to keep the pressure at $p\leq
10^{-3}\,\mathrm{mbar}$.\par

If access to the interior vacuum chamber is required, the thermal
reservoir and the attached collision volume can be heated up to
ambient temperature using 34 meters of heat rope with a maximum
power of $840\,\mathrm{W}$ which is wound around the copper
block.\par

\begin{figure}[!b]
        \begin{center}
            \includegraphics[width=1.0\columnwidth]{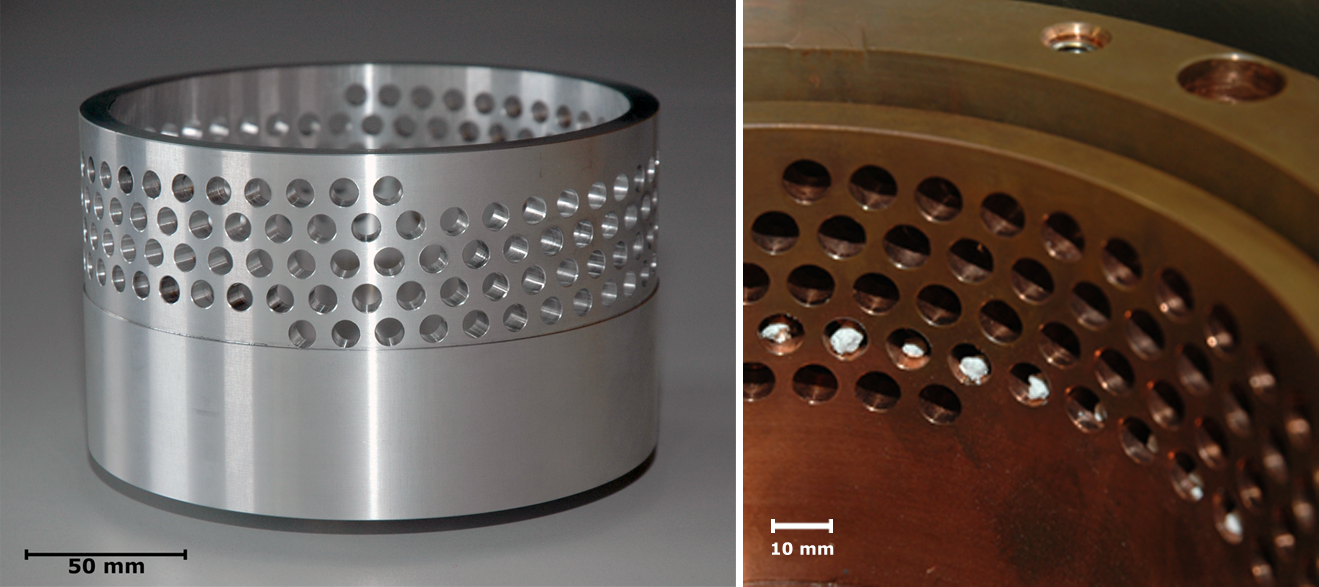}
            \caption{The cylindrical sample reservoir offers storage capacity for 180 individual millimeter-sized
            projectiles. On the left, the aluminum-made sample repository and to the right, the partly filled individual
            compartments of the copper reservoir.}
            \label{f_coliseum_alu}
        \end{center}
\end{figure}

On top of the thermal reservoir one of two cylindrical sample
repositories (Fig. \ref{f_coliseum_alu}) -- made from copper or
aluminum, respectively -- is attached to a brass-made quintuple
thread which is fit to a bore in the center of the copper block,
thereby establishing good thermal contact to the reservoir.

The sample carrier offers space for storing 180 projectile
aggregates which can be placed inside a double-helix of 90 holes
each. Thus, 90 collisions of pairs of aggregates or 180 impacts of
aggregates with a larger target can be performed by manually
rotating the cylindrical storage. In order to prevent the
projectiles from falling out of the storage compartments, and for
shielding the aggregates from radiative warming, a copper-made lid
covers the sample repository.\par

\subsection{\underline{Particle Acceleration Mechanism}}\label{ss_pistons}

To study collisions in the velocity range
$0.1\,\mathrm{m\,s^{-1}}\lesssim v_c \lesssim
0.5\,\mathrm{m\,s^{-1}}$, which is of astrophysical relevance
\citep{wei_cuzz1993,wei1997}, the aggregates are accelerated by a
set of synchronized, hydraulic pistons. The hydraulic actuators
are driven by an electrical DC motor at constant acceleration
level which ensures that the projectiles do not experience
accelerations exceeding $10\,g_0$. The two hydraulic pistons are
mounted opposite each other at the outside of the vacuum chamber,
and are guided to the sample repository passing through a
mechanical feedthrough, followed by a guiding
PTFE\footnoteT{\textbf{P}oly\textbf{t}etra\textbf{f}luoro\textbf{e}thylene}
piece, until they pick up the projectiles. To prevent the
aggregates from falling off the conical stainless steel rods they
are accelerated through guiding tubes of $80\,\mathrm{mm}$ length
until the pistons reach their outmost position, where switches are
activated to abruptly stop and rapidly retract them.\par

\subsection{\underline{Removable Collision Target}}\label{ss_target}

\begin{figure}[t!b]
        \begin{center}
            \includegraphics[width=1.0\columnwidth]{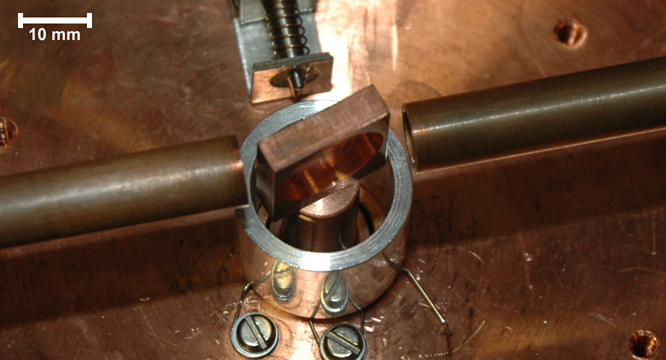}
            \caption{Pivot-mounted target frame in the collision volume (center). To the left and
            right of the target two cylinders are visible, within which the projectiles are
            accelerated from their storage compartments. In the center top of the image the target-release-mechanism can be seen, where the
            target can be lowered out of the collision volume during flight to allow for aggregate-aggregate collisions.}
            \label{f_target}
        \end{center}
\end{figure}

In addition to particle-particle collisions, a target screen (Fig.
\ref{f_target}), in which micron-sized dust grains can be packed
to simulate the surfaces of larger protoplanetary bodies, is used
to probe collisions of small particles with larger ones.

A mould of $17\,\mathrm{mm}$ diameter and $2\,\mathrm{mm}$ depth
is machined in either side of a copper frame, and installed
between the two guiding tubes at the center of the collision
volume and there attached to the sample repository. To vary the
impact angle, the target screen rotates by eight degrees per
collision (separation between neighboring storage compartments).
Thus, almost arbitrary impact angles in the range $0\textdegree -
75\textdegree$ (limited by the edges of the target frame blocking
the exits of the guiding tubes) can be achieved. Additionally, a
solenoid release mechanism offers the opportunity to drop the
target to a fixed position at the bottom of the collision volume
and thus, allows us to switch to particle-particle collisions.\par

\subsection{\underline{Imaging and Data Acquisition}}\label{ss_camera}

Recording of image sequences of the collisions is done by a
high-speed, high-resolution CMOS\footnoteT{\textbf{C}omplementary
\textbf{M}etal \textbf{O}xide \textbf{S}emiconductor} camera and a
high-performance recorder computer. The camera was operated at a
continuous accumulation of $107\,\mathrm{fps}$ of 8 bit grayscale
images of $1280\times1024\,\mathrm{px}$. Thus, the field-of-view
covers a plane of $24\times20\,\mathrm{mm^2}$ at a focal depth of
$\sim 5\,\mathrm{mm}$. The recording system is capable of writing
images at a maximum data rate of $133\,\mathrm{MByte\,s^{-1}}$ to
its 4 hard disks (totalling $260\,\mathrm{GByte}$), corresponding
to a maximum recording time of 33 minutes.\par

To capture the collisions the camera is mounted to a viewport at
the vacuum chamber's top flange. The collision volume is
illuminated by two synchronized stroboscopic (Xenon) flash lamps
which are also synchronized with the camera's image acquisition at
$107\,\mathrm{Hz}$.\par

\section{\underline{Microgravity Experiments}}\label{s_micro_g_exp}

During ESA's $45^{\mathrm{th}}$ Parabolic Flight Campaign
collision experiments were carried out at ambient temperatures
($\sim 300\,\mathrm{K}$) and a pressure of $2.8\cdot
10^{-1}\,\mathrm{mbar}$. Projectiles were prepared from
$2.5\,\mathrm{cm}$ RBD dust aggregates built from monodisperse
$1.5\,\mathrm{\mu m}$-sized $\mathrm{SiO_2}$ grains. The
$2-5\,\mathrm{mm}$ dust aggregates were cut from the larger `dust
cakes' using a razor blade. Recent x-ray tomography measurements
showed that at the cutting edges of the dust aggregates a slight
compaction from $\phi=0.15$ to $\phi \approx 0.17$ can be
observed.\par

\begin{figure}[t]
        \begin{center}
            \includegraphics[width=0.9\columnwidth]{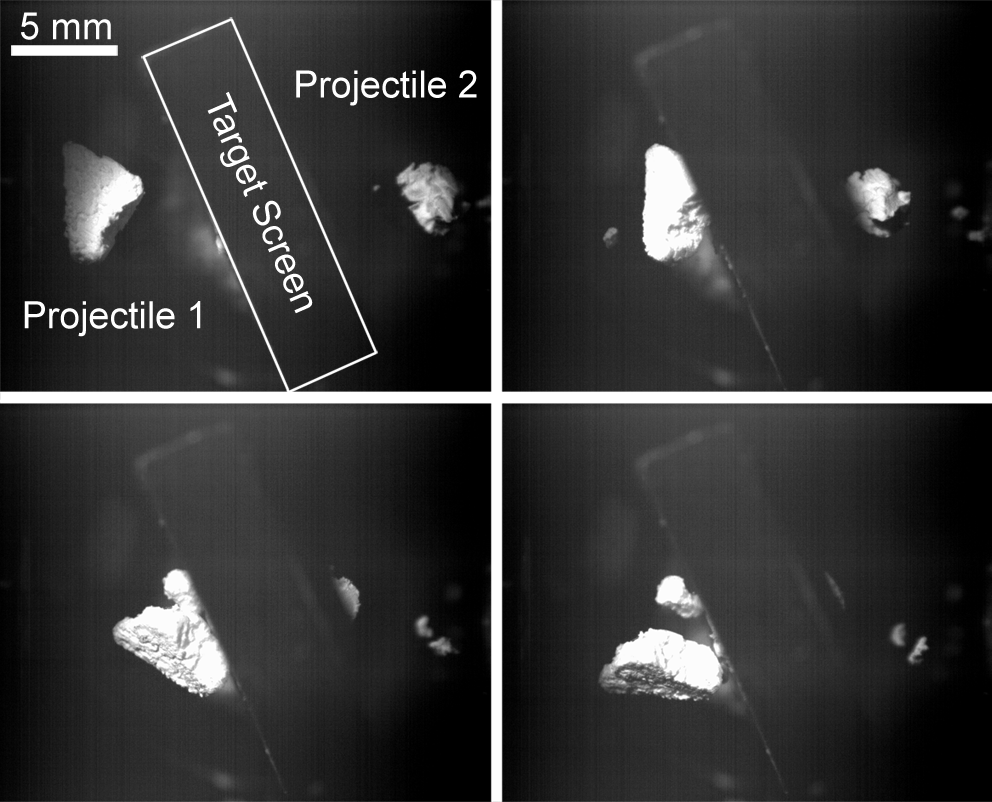}
            \caption{Example of a collision of a high-porosity dust aggregate with a compact dust
            target. The collision velocity is $\sim 0.2\,\mathrm{m\,s^{-1}}$. The time between subsequent
            images is $19\,\mathrm{ms}$.}
            \label{f_collage_pt4}
        \end{center}
\end{figure}

These projectiles were used to perform $3-4$ collisions per
parabola. Thus, images sequences of 76 impacts of dust aggregates
against a compact dust target of $\phi=0.24$ could be recorded
(see Fig. \ref{f_collage_pt4} for an example). It was observed
that in $\sim 90\%$ of the impacts the projectiles bounced off the
target, whereas $\sim 10\%$ of the collisions resulted in sticking
of -- only very small -- projectiles. Fragmentation was negligible
for this type of experiment.\par

\begin{figure}[t]
        \begin{center}
            \includegraphics[width=0.9\columnwidth]{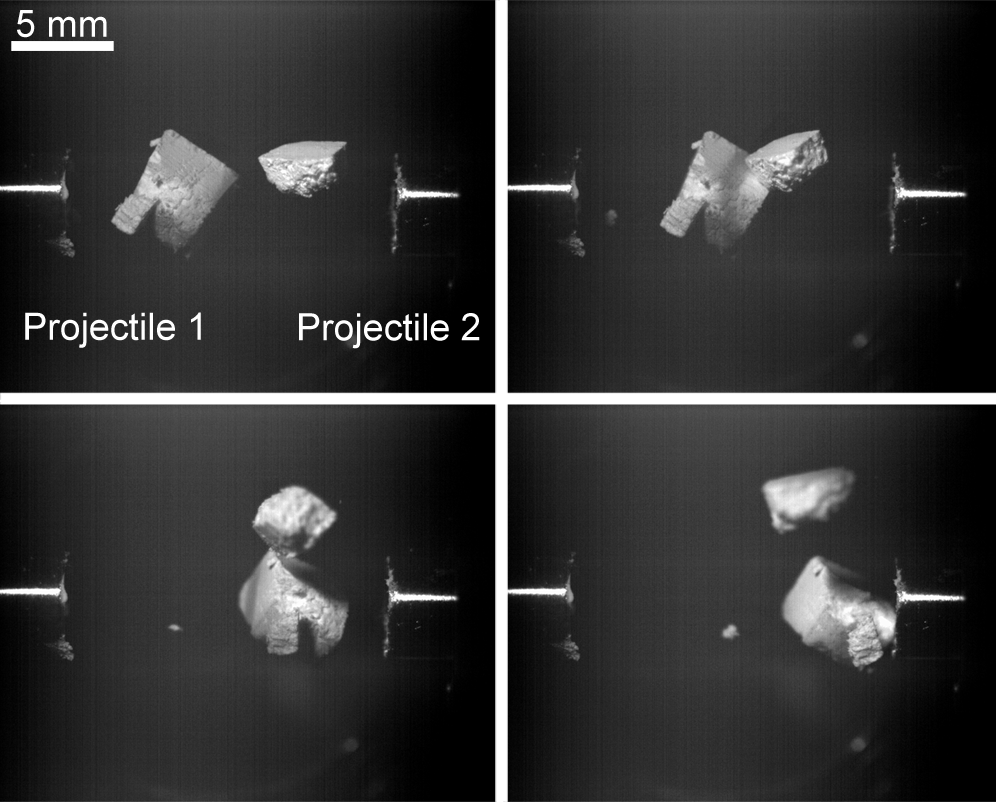}
            \caption{Example of a collision between two high-porosity dust aggregates.
            The collision velocity is $\sim 0.4\,\mathrm{m\,s^{-1}}$. The time between subsequent
            images is $19\,\mathrm{ms}$.}
            \label{f_collage_pp4}
        \end{center}
\end{figure}

In addition to the particle-target collisions, 39 encounters of
pairs of fluffy dust aggregates were captured (see Fig.
\ref{f_collage_pp4} for an example). Analysis of the image
sequences showed that again $\sim 90\%$ of the collisions were
dominated by rebounding behavior. For this type of experiment
fragmentation was observed in $\sim 10\%$ of the collision events,
whereas in this case sticking was unimportant.\par

Analysis of the relative velocities of the colliding bodies before
and after the impact, calculation of the coefficient of
restitution and determination of the normalized impact parameter
(for aggregate-aggregate collisions) and impact angle (for
aggregate-target collisions), respectively, show that for a
central collision only $\sim 5\%$ of the translational energy is
conserved. This value increases with increasing normalized impact
parameter. Additionally, from the image sequences it becomes clear
that a significant amount of energy is transformed into rotational
motion of the aggregates after the encounter. A quantitative
analysis could not be performed, because the narrow depth of
field, and thus, blurring effects ruled out the determination of
the rotational velocity and the irregular shape of the aggregates
made a determination of the moment of inertia impossible.\par

\section{\underline{Experiment Improvements}\\\underline{for Future Campaigns}}\label{s_improvements}

To improve the experimental setup and to increase the quality of
the obtained data for future Parabolic Flight Campaigns a number
of changes will be or were already made. During ESA's
$45^{\mathrm{th}}$ PFC only the copper-made sample repository was
used. In order to save valuable time during preparation of the
parabolic flights a second, aluminum-made sample reservoir was
machined. This can be prepared for the upcoming flight by the
ground crew, while the flight crew uses the other one to perform
microgravity experiments. Thus, the optimized procedures are
expected to save $2.5$ hours of preparation time between
subsequent flights.\par

Additionally, to increase the quality of the recorded image
sequences and remove any shadowing effects the stroboscopic
illumination will be modified by mounting two synchronized flash
lamps at an angle of $\alpha = 45\textdegree$ to the pistons' axis
and at an angle of $\beta = 56\textdegree$ to the camera axis. The
high-speed camera will be extended by a beam splitter optics which
allows to record image sequences with an angular separation of
$\psi = 60\textdegree$. Thereby, three-dimensional collision
information -- allowing the exact determination of the impact
parameter -- can be obtained.\par

To optimize the separating of the projectile from the pistons'
conical heads the hydraulic acceleration mechanism will be
replaced by a synchronized master-and-slave system of electric DC
motors. In contrast to the previously used hydraulic ones, these
motors with their attached leadscrews will be equipped with a
hard, mechanical stop confirming the instantaneous deceleration of
the pistons' rods allowing a more accurate separation of the
aggregates.\par

\section*{\underline{Acknowledgments}}\addcontentsline{toc}{chapter}{Acknowledgments}

We thank J. Barrie (\strath), G. Borst (Dutch Space), M. Costes
(University of Bordeaux I), R. W. Dawson (\strath), G. Drinkwater
(\strath), F. Gai (Novespace), K. Gebauer (\tubs), J. Gillan
(\strath), M. Hutcheon (\strath), E. Jelting (\tubs), L. Juurlink
(\leiden), M. Krause (\tubs), E. de Kuyper (\leiden), H. Linnartz
(\leiden), J. Lub (\leiden), F. Molster (NOVA), A. Orr (ESA), D.
Salter (\leiden), B. Stoll (\tubs), P. Tuijn (\leiden), G. van der
Wolk (\rug) and the members of the ESA Topical Team:
Physico-Chemistry of Ices in Space, ESTEC Contract No:
15266/01/NL/JS.\par

The project was generously supported by Air Liquide, Kayser-Threde
GmbH, Pfeiffer Vacuum, the Radboud University Nijmegen and VTS
Ltd.\par

The ICES experiment was co-funded by the German Space Agency (DLR)
under grant No. 50 WM 0636, the European Space Agency (ESA), the
Leids Kerkhoven-Bosscha Fonds (LKBF), the Netherlands Research
School for Astronomy (NOVA), the Netherlands Institute for Space
Research (SRON) under grant No. PB-06/053 and the Scottish
Universities Physics Alliance (SUPA) Astrobiology Equipment
Fund.\par

We are indebted to the late Ron Huijser for many stimulating
discussions.

\newcommand{\mnras}{Monthly Notices of the Royal Astronomical Society}
\newcommand{\prl}{Physical Review Letters}
\newcommand{\apj}{The Astrophysical Journal}
\newcommand{\rsi}{Review of Scientific Instruments}
\newcommand{\jcp}{The Journal of Chemical Physics}
\newcommand{\nat}{Nature}
\addcontentsline{toc}{chapter}{References}%
\lhead[\thepage]{\nouppercase{}}%
\rhead[\nouppercase{}]{\thepage}%
\bibliography{literature}%

\end{document}